\pdfoutput=1
\documentclass[11pt]{article}
\usepackage{jheppub}

\usepackage[utf8]{inputenc}

\usepackage{amsmath}
\usepackage{amssymb}
\usepackage{graphicx,color}
\usepackage{amsthm}
\usepackage{mathrsfs}

\usepackage{tikz}
\usetikzlibrary{matrix,arrows}

\usepackage{ifpdf}

\usepackage{subfigure}

\usepackage[boxsize=0.5em,aligntableaux=center]{ytableau}

\usepackage{tikz}
\usetikzlibrary{shapes,shadows,calc}
\usepgflibrary{arrows}
\usepackage{rotating}
\usepackage{wrapfig}

\tikzset{
  sshadow/.style={opacity=.25, shadow xshift=0.05, shadow yshift=-0.06},
}

\def\kbbox[#1,#2,#3,#4,#5]#6{
        \draw[dashed] node[draw,color=gray!50,minimum
        height=#1,minimum width=#2] (#4) at #5 {}; 
        \node[anchor=#3,inner sep=2pt] at (#4.#3)  {#6};
}
\def\kbboxred[#1,#2,#3,#4,#5]#6{
        \draw[] node[draw,color=red,minimum
        height=#1,minimum width=#2] (#4) at #5 {}; 
        \node[anchor=#3,inner sep=2pt] at (#4.#3)  {#6};
}

\newcommand{\rep}[1]{\mathbf{#1}}

\newcommand{\bP}{\mathbb{P}}

\newcommand{\bZ}{\mathbb{Z}}

\newcommand{\cL}{\mathcal{L}}

\newcommand{\cN}{\mathcal{N}}

\newcommand{\cV}{\mathcal{V}}

\newcommand{\tX}{\mathcal{X}}

\newcommand{\hX}{\mathbb{X}}

\def\Label#1{\label{#1}%
  \smash{\hbox to0pt{\raise1ex\hbox{\tiny[#1]}\hss}}}
\def\noLabels{\let\Label=\label}
\def\nobbibitem{\let\bbibitem=\bibitem}
 \def\noBibitem{\let\Bibitem=\bibitem}

\newcommand{\be}{\begin{equation}}
\newcommand{\ee}{\end{equation}}
\newcommand{\beq}{\begin{equation}}
\newcommand{\eeq}{\end{equation}}
\newcommand{\bea}{\begin{eqnarray}}
\newcommand{\eea}{\end{eqnarray}}
\newcommand{\R}{\text{Re}}
\newcommand{\I}{\text{Im}}

\newcommand{\ba}{\begin{eqnarray}}
\newcommand{\ea}{\end{eqnarray}}

\newcommand\varpm{\mathbin{\vcenter{\hbox{%
  \oalign{\hfil$\scriptstyle+$\hfil\cr
          \noalign{\kern-.3ex}
          $\scriptscriptstyle({-})$\cr}%
}}}}
\newcommand\varmp{\mathbin{\vcenter{\hbox{%
  \oalign{$\scriptstyle({+})$\cr
          \noalign{\kern-.3ex}
          \hfil$\scriptscriptstyle-$\hfil\cr}%
}}}}

\title{\centering Yukawas and discrete symmetries in\\F-theory
  compactifications without section}

\author[a]{I\~naki Garc\'ia-Etxebarria,}
\author[a]{Thomas W. Grimm,}
\author[a]{and Jan Keitel}

\affiliation[a]{Max Planck Institute for Physics,\\
F\"ohringer Ring 6, 80805 Munich, Germany}

\emailAdd{inaki@mpp.mpg.de}
\emailAdd{grimm@mpp.mpg.de}
\emailAdd{jkeitel@mpp.mpg.de}

\abstract{In the case of F-theory compactifications on genus-one
  fibrations without section there are naturally appearing discrete
  symmetries, which we argue to be associated to geometrically
  massive $U(1)$ gauge symmetries. These discrete
  symmetries are shown to induce non-trivial selection rules for the allowed Yukawa
  couplings in $SU(N)$ gauge theories. The general discussion is exemplified using a
  concrete Calabi-Yau fourfold realizing an $SU(5)$ GUT model. We
  observe that M2 instanton effects appear to play a key role in the 
  generation of new superpotential terms and in the dynamics close to phase transition loci.}

\begin{document}
\setlength{\parskip}{5pt}

\makeatletter
\let\old@fpheader\@fpheader
\renewcommand{\@fpheader}{\old@fpheader\hfill
MPP-2014-333}
\makeatother

\maketitle
\newpage

\section{Introduction}

F-theory \cite{Vafa:1996xn} compactifications to four dimensions are
typically defined by specifying a $T^2$ fibered Calabi-Yau
fourfold. The traditional assumption is that the fibration has a
section, i.e. there is an embedding of the basis divisor into the
total space, almost everywhere intersecting the fiber at a point. All
such models are birational to a Weierstrass model
\cite{Nakayama}. %and we will give a (highly selective) of the work done on them in the following. 
Restricting oneself to Calabi-Yau fourfolds defined
by Weierstrass models (and thus, having at least one section)
simplifies model building with non-Abelian gauge symmetries, since
there are well understood techniques for reading off the low energy
non-abelian gauge groups from the structure of a Weierstrass model\footnote{See for example Table 4 of \cite{Grassi:2011hq}
for a comprehensive dictionary between vanishing degrees of the Weierstrass model
and the associated gauge algebras.}.
Considerable effort has been made to develop
similar techniques for analyzing and engineering elliptically fibered
Calabi-Yau manifolds that also give rise to Abelian gauge groups in the
low energy effective theory.

Initiated by the construction of the $U(1)$-restricted model
in \cite{Grimm:2010ez}, the study of global F-theory compactifications with
$U(1)$ gauge factors can very roughly be divided into two approaches:
(1) For a given $U(1)$ gauge rank, one can determine the ambient space in which
every elliptic fiber giving rise to such a low energy theory must be embeddable by 
using an old idea of Deligne \cite{deligne1975courbes}. Having obtained this space,
one can then try to extract information about generic features of all such compactifications,
such as all the matter representations can that possibly occur \cite{Morrison:2012ei,Borchmann:2013jwa,
Cvetic:2013nia, Borchmann:2013hta, Cvetic:2013jta,Cvetic:2013qsa}.
Non-generic elliptic fibers in Tate form were studied in \cite{Mayrhofer:2012zy, Kuntzler:2014ila}.
(2) Conversely, one can take the stand and demand that given an arbitrary elliptically
fibered Calabi-Yau manifold, one should be able to determine the low energy effective
theory it gives rise to \cite{Braun:2013yti, Grimm:2013oga, Braun:2013nqa}. By breaking up
the Calabi-Yau into its various building blocks and determining
which of them can be treated separately, one can then systematically answer questions about entire
classes of compactification manifolds \cite{Braun:2013nqa} and find connections between them\nocite{Braun:2014oya,Morrison:2014era,Anderson:2014yva}
using Higgsings \cite{Klevers:2014bqa}. Alternatively, one could perform computer-aided scans
over large numbers of compactifications as was done for example in \cite{Braun:2011ux,Krippendorf:2014xba}.
Naturally, these two approaches are not mutually exclusive and there exist many ways in
which they overlap. Additionally, work has been done to understand the geometry
associated to singularities in higher codimensions in the base manifold \cite{Lawrie:2012gg} and
the relations between the different ways of resolving these
\cite{Hayashi:2013lra, Hayashi:2014kca, Braun:2014kla}. Furthermore, we note that a perpendicular
approach has been taken by \cite{Grassi:2013kha, Grassi:2014sda},
where resolutions are avoided by instead deforming the singularities,
corresponding to a Higgsing of the gauge group.

By now, not only the Abelian gauge groups themselves, but also purely Abelian matter states, 
often called \emph{singlet states}, appear to be fairly well understood in four 
and six dimensions, both from a geometric \cite{Krause:2011xj, Cvetic:2013nia, Cvetic:2013uta} 
and a field theoretic perspective \cite{Grimm:2011fx, Cvetic:2012xn, Grimm:2013oga} using the 
Chern-Simons terms of the effective theory compactified on a circle. Recently, a proposal for
counting the precise number of multiplets in F-theory compactifications to four 
dimensions has been made \cite{Bies:2014sra}.
In contrast, Yukawa couplings in global compactifications have been much less studied so 
far, both those that involve singlets and those that do not. While their assumed geometrical 
counterparts, intersections of different matter curves in codimension $3$ in the base manifold, 
have received attention \cite{Mayrhofer:2012zy, Borchmann:2013jwa, Cvetic:2013nia, Borchmann:2013hta,
Cvetic:2013jta, Cvetic:2013uta, Lin:2014qga, Klevers:2014bqa}, it appears crucial to point out that
the relation to T-branes \cite{Cecotti:2010bp, Anderson:2013rka}, and in particular the
low energy effective theory and local 
models \cite{Hayashi:2009ge,Font:2009gq,Aparicio:2011jx,Marsano:2011hv,Font:2012wq,Font:2013ida}
 remain to be explored.

Notably, beyond mathematical convenience there is no a priori physical
reason to restrict oneself to $T^2$ fibrations with
section. Calabi-Yau fourfolds with $T^2$ fiber but no section
constitute perfectly respectable M-theory backgrounds, and they can
admit a F-theory limit. The physics of such backgrounds is rather
interesting, and only recently it has been started to be
systematically explored, mostly for the case of compactifications on
threefolds
\cite{Braun:2014oya,Morrison:2014era,Anderson:2014yva}.\footnote{See
  also \cite{Berglund:1998rq,deBoer:2001px} for earlier work on the topic.}
 In
this paper we extend the physical picture put forward in \cite{Anderson:2014yva}
to Calabi-Yau fourfold compactifications without section. 
We propose a closed string and an open string perspective on 
the massive $U(1)$ arising in compactifications without section, 
and comment on the geometrical configurations realizing this 
duality. Furthermore, we explicitly describe how a non-Abelian 
gauge theory on seven-branes can be engineered in 
such geometries. This allows us to argue that models without
section do have potentially fruitful model building properties, such
as the natural appearance of certain discrete symmetries at low
energies. These discrete symmetries can (and do) forbid certain Yukawa
couplings from being generated, even though the Yukawa couplings are
otherwise allowed by all continuous symmetries present at low
energies. 
Let us remark that intersecting D6 brane scenarios with similar
physical implications have recently been studied for example
in \cite{BerasaluceGonzalez:2011wy, Anastasopoulos:2012zu,Honecker:2013hda}.
As we were completing this paper,  \cite{Klevers:2014bqa} appeared 
in which discrete symmetries in F-theory compactifications
are also studied.

We start in section~\ref{sec:review} with a quick review of some
aspects of the physics of compactifications without section, where we
explain the connection of discrete symmetries to certain geometrically
massive $U(1)$ symmetries, and we highlight the relevance of including
M2 instanton effects in order to have a consistent description of the
physics. In section~\ref{sec:examples} we then provide a detailed
analysis of a phenomenologically motivated toy example, and show that
indeed discrete symmetries forbid certain Yukawa couplings from
appearing, in agreement with what the general discussion suggests.

\section{F-theory compactifications without section and Yukawa structures}
\label{sec:review}

In this section we first discuss F-theory on genus-one fibrations 
without section generalizing the insights of \cite{Anderson:2014yva}
to Calabi-Yau fourfold compactification. This will be the first task of
subsection \ref{Physics_without_section}, where we will provide 
two different perspectives, a closed string and an open string one, 
on massive $U(1)$ gauge symmetries arising from such fibrations. 
Next, we examine the discrete symmetries that survive as remnants
of the massive $U(1)$ gauge symmetries at low energies in section
\ref{ss:discrete}. The Yukawa structures that arise in
four-dimensional $SU(5)$ Grand Unified Theories are 
treated in subsection \ref{Yukawa-Structures}, putting special
emphasis on continuous and discrete selection rules. We also argue that an interesting 
class of instanton effects plays a key role in connecting the 
closed and the open string pictures. Finally, in subsection \ref{Higgstransition}
we give a more detailed geometric description of the set-up 
and discuss the string interpretation of the Higgsing. 

\subsection{Physics of F-theory compactifications without section} \label{Physics_without_section}

In this section we first review the physics of F-theory compactifications on
manifolds without section following the point of view taken in
\cite{Anderson:2014yva} (see also
\cite{Braun:2014oya,Morrison:2014era,Klevers:2014bqa}). Next, it will be
crucial to extend the discussion to a four-dimensional context,
i.e.~to the study of Calabi-Yau fourfolds without section.

Before turning to geometries without section, it is useful to first recall 
some facts about geometries with a section. In order that F-theory 
is well-defined, a potential Calabi-Yau compactification geometry should admit
a genus-one fibration over some base manifold $B$. 
In this case the value of the dilaton-axion $\tau$, given 
by the complex structure modulus of the $T^2$ fiber, can be extracted from the geometry 
and describes a Type IIB string theory background.
A subclass of such $T^2$ fibrations are geometries that have 
a section. The presence of a section implies the existence of a global
meromorphic embedding of the base $B$ into the total space of the fibration.
Alternatively, one can view a section as selecting precisely one point in the fiber over
every point in the base with the possible exception of
lower-dimensional loci in the base where the fiber degenerates. 
Fibrations with a section can be birationally transformed into a Weierstrass model 
given by 
\begin{align} \label{eq:Weierstrass}
y^2 = x^3 + f x z^4 + g z^6 \ ,
\end{align}
where $(x: y: z)$  are the homogeneous coordinates of a $\mathbb{P}^{2,3,1}$, and $f,g$ are 
functions on $B$. A canonical section is simply given by $z = 0$. While the F-theory literature
has mostly focused on such Weierstrass models, the 
presence of a section is by no means a physical requirement for the 
existence of an effective F-theory action. 

Let us thus turn to the case of having a genus-one fibered Calabi-Yau fourfold $\tX_4$ 
without section. Despite the absence of a section such geometries still admit a 
multi-section or $n$-section \cite{Braun:2014oya,Morrison:2014era,Anderson:2014yva,Klevers:2014bqa}.
More precisely, while one cannot find a divisor cutting out a single point in the fiber
over every point in the base, one \emph{can} still find divisors singling out $n$ 
points in the fiber. These points may then undergo monodromies as one moves along
the base  $B$ of $\tX_4$.
 The simplest case, which will be also the main focus in this work, 
is the situation where $n=2$, i.e.~a manifold with a bi-section. It was
argued in \cite{Anderson:2014yva} 
that the effective action of F-theory compactified on such a manifold should include 
a massive $U(1)$ gauge symmetry. In fact, one should rather think of 
the compactification as a cousin of a reduction with two sections, which 
would correspond to having an extra $U(1)$ gauge symmetry present in the effective theory. 
Since the $U(1)$ is massive in compactifications without section, let us 
recall that a $U(1)$ can become massive by two related mechanisms: 
a linear Higgs mechanism or a non-linear Higgs mechanism, also 
known as the St\"uckelberg mechanism. It was argued in \cite{Anderson:2014yva} 
that both points of view are useful to specify the effective theory 
obtained from a $\tX_4$ compactification. 

We start by describing the St\"uckelberg picture first. In this case 
the F-theory effective theory on $\tX_4$ contains a $U(1)$ that is massive
due to the shift-gauging of an axion $c$ given by
\begin{align}  \label{Stuckgauging}
   \mathcal{\hat D} c =  d c + m \hat A^1\ ,     \qquad c \rightarrow c - m \Lambda\ .
\end{align}
Upon `eating' the axionic degrees of freedom the kinetic term of $c$ turns into 
a mass term for $\hat A^1$. It was argued in  \cite{Anderson:2014yva} that for a geometry without 
section the axion involved in the gauging is a closed-string degree of freedom 
arising from the R-R or NS-NS two-form of Type IIB string theory. In other words, 
the geometries realize a geometrically massive $U(1)$ gauge symmetry \cite{Grimm:2010ez,Grimm:2011tb,Braun:2014nva}.
In fact, at weak string coupling $c$ is simply the zero-mode of the R-R two-form $C_2$
that renders a D7-brane $U(1)$ massive \cite{Jockers:2004yj}. 

Let us briefly recall the argument to justify that F-theory compactifications 
with a bi-section do indeed yield a St\"uckelberg massive $U(1)$ in the effective theory. 
Following the suggestion of \cite{Witten:1996bn}, it was shown in \cite{Anderson:2014yva}
that the M-theory to F-theory duality for such geometries requires the introduction of
a background flux on the F-theory side. 
In order to connect M-theory and F-theory one has to consider the F-theory setup on an extra circle. 
Following the duality, the absence of a section requires to introduce circle flux $n$ along the extra circle.
Indeed, at weak coupling this is due to a background flux for the field strength
of the R-R two-form $C_2$. In the lower-dimensional 
theory the circle flux induces a further gauging 
\beq \label{complete_gauging}
   \mathcal{ D} c =  d c + m  A^1 + n A^0\ ,
\eeq
where $A^0$ 
is the Kaluza-Klein vector. Taking into account that this implies the presence of a St\"uckelberg mass
for the $U(1)$ gauge field given by the linear combination $m  A^1 + n A^0$, 
it was shown that the M-theory and F-theory effective theories 
can indeed be matched. 
The presence of the St\"uckelberg gauging \eqref{Stuckgauging} coupling to the R-R or NS-NS two-form 
axion is crucial for this match to work. 

As pointed out above, the study of F-theory compactifications without section has so far
focused on Calabi-Yau threefolds \cite{Braun:2014oya,Morrison:2014era,Anderson:2014yva,Klevers:2014bqa}.
However, it is important to remark on how these considerations generalize to four-dimensional 
F-theory compactifications on Calabi-Yau fourfolds.  In a four-dimensional theory 
with $\cN=1$ supersymmetry the axion $c$ must arise from a complex field. 
We take it to be the real part of a complex field $G$, $\R\, G = c$. 
The field $G$ is obtained when expanding the M-theory three-form as \cite{Haack:1999zv,Grimm:2010ks}
\begin{align}
   C_3 = i G \bar \Psi - i \bar G \Psi\ ,
\end{align}
where $\Psi$ is a $(2,1)$-form on the Calabi-Yau fourfold $\tX_4$. Using 
this definition of $G$, one can derive the four-dimensional effective 
theory. The relevant $U(1)$ gauging appears in the kinetic term of $G$ given by 
\begin{align} \label{kinterm}
   \cL_4 = K_{G \bar G} \mathcal{\hat D} _\mu G\, \mathcal{\hat D} ^\mu \bar G\ , \qquad    \mathcal{\hat D}  G = d G + m \hat A^{1}\ .   
\end{align}
Upon `eating' the axion $\R\, G$, the kinetic term \eqref{kinterm} becomes a mass term for $\hat A^1$, 
and the mass is simply given by $K_{G\bar G}$.
%Note that due to $\cN=1$ supersymmetry the metric $K_{G \bar G}$ is the second derivative 
%of a K\"ahler potential. 
Furthermore, it was shown in \cite{Haack:1999zv,Grimm:2010ks} that for a massless
$G$ $K_{G \bar G}$ takes the form 
\begin{align}
    K_{G \bar G} = \frac{i}{2 \cV} \int_{\tX_4} J \wedge \bar \Psi \wedge \Psi\ .
\end{align}
Note that since $\Psi$ is a $(2,1)$-form on $\tX_4$, it depends on the complex structure moduli $z^k$ of 
$\tX_4$. Remarkably, the moduli dependence of $\Psi$ can be specified by a \textit{holomorphic} function $h(z)$.
In the simplest situation one finds that \cite{Grimm:2010ks,Grimm:2014vva}
\begin{align}
    K_{G \bar G} \ \propto \ (\I h)^{-1}\ .
\end{align}
Moving along the complex structure moduli space, the coupling $K_{G \bar G}$ setting the mass of the $U(1)$
can become zero. 

Let us comment on the points at which the $U(1)$ becomes massless.
In order to do that, we extrapolate the behavior of 
$K_{G \bar G}$ using the results from a Calabi-Yau threefold. Indeed,
the analog coupling in a Calabi-Yau threefold compactification depends crucially 
on the complex structure moduli and can be specified by a holomorphic pre-potential $\mathcal{F}(z)$. 
In this case, the function $h$ can be thought of as a second derivative of the pre-potential $\mathcal{F}(z)$.
One then expects 
that at special points $z^i \approx 0, \ i =1,\ldots, n_{\rm con}$ in complex structure moduli space one has 
\begin{align}
   h(z) = \sum_i \, a_i \log z^i  + \ldots  \ ,
\end{align}
where $a_i$ are constants and the dots indicate terms that are polynomial in the complex structure parameter $z^i$.
Geometrically, as we discuss in more detail below, this indicates that  
 the points $z^i=0$ are conifold points and a geometric transition takes place. 
In fact, as discussed already in \cite{Anderson:2014yva}, the Calabi-Yau threefold with a bi-section $\tX_3$ can 
transition to a Calabi-Yau threefold with two sections $\hX_3$ by means of a conifold transition. 
In the Calabi-Yau fourfold case a similar transition from $\tX_4$ to $\hX_4$ can 
take place. In this case, however, one finds a whole curve of conifold points:
\begin{align}
    \tX_3 & \quad \xrightarrow{\ \ \text{tune}\ z^i\ \ }  \quad  \tX_3^{\rm sing} \ \text{with conifold points}\quad \xrightarrow{\ \ \text{resolve} \ \ } \quad  \hX_{3} \\
     \tX_4 & \quad \xrightarrow{\ \ \text{tune}\ z^i\ \ }  \quad  \tX_4^{\rm sing} \ \text{with conifold curve}\quad \xrightarrow{\ \ \text{resolve} \ \ } \quad  \hX_{4}
\end{align}
We stress that the resolved branch $\hX$ can only be accessed in 
the lower-dimensional theory, i.e.~in M-theory on $\hX$. Nevertheless, the 
existence of the branch $\hX$ naturally leads us to another interpretation 
of the setup with a $U(1)$ made massive by a linear Higgs mechanism.

To introduce the linear Higgs mechanism picture, let us approach the singular 
geometry from the side of $\hX_4$. At the singular point one also finds that 
there are new matter states in the four-dimensional effective theory, that 
are charged under the $U(1)$. In other words, these admit the couplings 
\begin{align}
     \mathcal{\hat D}  \phi = d \phi +  i \hat q \hat A^{1} \phi\ ,
\end{align}
where $\hat{q}$ is the $U(1)$ charge of the complex field $\phi$. This implies 
that one can also think of giving a mass to the $U(1)$ by turning on 
a vacuum expectation value (VEV) for the field $\phi$. In the 
F-theory compactifications under consideration the field $\phi$ will 
be a matter field arising from the open string sector on intersecting 
seven-branes. It will further be a singlet under the any additional non-Abelian group
and therefore denoted by ${\bf 1}_{\hat q}$, where the subscript indicates
the $U(1)$-charge. Working with the open string matter field $\phi$ should be considered 
as the dual picture to working with the closed string field $G$. In order to match the 
charges one expects an identification 
\begin{equation} \label{open-closed-map}
   {\bf 1}_{\hat q} \ \text{(open string)} \quad \leftrightarrow \quad A(z) e^{2\pi i r G}\ \text{(closed string)}  \ , 
\end{equation}
where $m r = - \hat q $, and $A(z)$ is a coefficient that generally depends 
on the complex structure moduli of $\tX_4$. Working
with either ${\bf 1}_{\hat q}$ or $G$ degrees of freedom should give a dual description 
of the same physical effective theory.

Let us close this section by noting that the fact that the $U(1)$ is massive implies
that it will be absent in the effective theory at energy scales below its mass.
In this effective theory the selection rules originally imposed by the $U(1)$ gauge symmetry will 
remain as discrete symmetries. In the next section we therefore discuss discrete gauge symmetries
of F-theory compactifications in more detail.

%We will discuss the symmetries of the 
%Yukawa couplings of an $SU(5)$ GUT realized in such a geometry 
%next.

\subsection{Discrete gauge symmetries} \label{ss:discrete}
Let us now examine
the Higgsing with respect to the discrete symmetries left over and use the restrictions
that general F-theory spectra have to fulfill to our advantage. The set-up
we consider consists of a $U(1)_{0} \times U(1)_{1} \times SU(N)$ gauge group in the
circle-compactified theory
with matter states in the singlet, the fundamental $\rep{N}$ and the antisymmetric 
representation $\rep{\frac{1}{2}N (N-1)}$
of $SU(N)$. Here $U(1)_{0}$ is the gauge group corresponding to the 
Kaluza-Klein vector and $U(1)_{1}$ is the gauge factor lifting to the proper four-dimensional $U(1)$ factor. Hence they correspond
to the gauge fields $A^0$ and $A^1$ in \eqref{complete_gauging} of the previous section.
As noted for example in \cite{Braun:2013yti, Braun:2013nqa}, the allowed $U(1)_{1}$ charges of all the occurring representations obey certain restrictions. First of all, let us assume that the $U(1)_{1}$ generator has been rescaled such that the smallest singlet charge is $N$, thereby ensuring that all there are no fractional charges under $U(1)_{1}$. Then the $U(1)_{1}$
charges of the matter states fundamental representation satisfy
\begin{align} \label{eq:split_fund}
Q_{U(1)_{1}}(\rep{N}) \equiv k \mod N
\end{align}
and the charges of the states in the antisymmetric representations fulfill
\begin{align} \label{eq:split_anti}
Q_{U(1)_{1}}\left(\rep{\tfrac{1}{2}N (N-1)}\right) \equiv 2k \mod N\,,
\end{align}
where $k$ is an integer defining the \emph{matter split} \cite{Braun:2013nqa} with respect to $U(1)_{1}$.
Let us now assume that a field in the $\rep{1}_{m, \hat{n} N}$ representation\footnote{That is, the field transforms
trivially under $SU(N)$ and has charges $m$ and $\hat{n} \cdot N$ under $U(1)_{0}$ and $U(1)_{1}$, respectively.
Note that $m$ here is the same as in \eqref{complete_gauging}, but
we have chosen to write split up the $n$ from \eqref{complete_gauging} into $n = \hat{n} \cdot N$ in order to
emphasize that it contains a factor of $N$.} attains a VEV. For a general spectrum, the $U(1)_{0} \times U(1)_{1}$ symmetry is broken
to $U(1)' \times (\mathbb{Z}_{m} \oplus \mathbb{Z}_{\hat{n} N})$ while the $SU(N)$ factor remains intact.
Here $U(1)'$ is the linear combination $\hat{n} N U(1)_{0} - m U(1)_{1}$ under which the singlet
with non-trivial VEV is uncharged. In terms of the old charges, 
the charges under the new gauge group are
\begin{align}
Q_{U(1)'} &= \hat{n} N Q_{U(1)_{0}} - m Q_{U(1)_{1}}
\end{align}
and
\begin{align}
Q_{\mathbb{Z}_m} &= Q_{U(1)_{0}} \mod m &
Q_{\mathbb{Z}_{\hat{n} N}} &= Q_{U(1)_{1}} \mod \hat{n} N\,.
\end{align}

Now let us be more specific and assume that the field Higgsing the $U(1)$ gauge factor has charges $m=1$ and $\hat{n}=2$, as we find
to be the case for all the models in which the elliptic fiber is embedded inside $\mathbb{P}_{112}$. Roughly speaking, this is due
to the fact that states that are doubly charged under $U(1)_1$ are intrinsically linked to states with non-trivial KK-charge,
since the zero section also appears as a term in the divisor acting as the four-dimensional $U(1)$ generator
\cite{Anderson:2014yva}, and it is these states that appear at the conifold singularities.
 At first sight, the discrete gauge symmetry then appears to be $\mathbb{Z}_{2 N}$. However, we argue that it is in fact only
 $\mathbb{Z}_2$. To see this, look at all the possible charges of the matter representations:
\begin{align}
Q_{\mathbb{Z}_{2N}} (\rep{1}) &\in \{0, N\}\,, \nonumber \\
Q_{\mathbb{Z}_{2N}} (\rep{N}) &\in \{k, k+N\}\,, &
Q_{\mathbb{Z}_{2N}} \left(\rep{\tfrac{1}{2}N (N-1)}\right) &\in \{2k, 2k+N\}\ .
\end{align}
Here we represent elements of $\mathbb{Z}_{2N}$ by integers and the group law by addition modulo $2N$.
This implies
\begin{align}
2 \cdot Q_{\mathbb{Z}_{2N}} (\rep{1}) &=0\,, \nonumber \\
2 \cdot Q_{\mathbb{Z}_{2N}} (\rep{N}) &=2k\,, &
2 \cdot Q_{\mathbb{Z}_{2N}} \left(\rep{\tfrac{1}{2}N (N-1)}\right) &= 4k\,,
\end{align}
which is an element of the center $\mathbb{Z}_N$ of the unbroken $SU(N)$ factor. We therefore see that we can
split $\mathbb{Z}_{2N}$ into $\mathbb{Z}_2 \oplus \mathbb{Z}_N$ and identify the second part with the
center of the non-Abelian gauge group.

Finally, let us note that there are at least two kinds of special cases for which our discussion has to be adjusted.
The first such case corresponds to a $0$-split, i.e.~spectra of the above type with $k=0$. In this case
all $U(1)$ charges are divisible by $N$ and the second part of the discrete gauge group
is $\mathbb{Z}_{\hat{n}}$ instead of $\mathbb{Z}_{\hat{n} N}$. Therefore the center of the $SU(N)$ group drops out directly.
The second case corresponds to set-ups where $N$ is even and $k = \frac{N}{2}$.
In that case there may be additional identifications because fields that we treated
independently above may be contained in the same multiplet.

\subsection{Four-dimensional Yukawa structures} \label{Yukawa-Structures}

In the following we discuss the Yukawa structures of $SU(5)$ GUTs
engineered in an F-theory compactification without section. Therefore,
let us consider a $SU(5)$ GUT with ${\bf 10}$ representations and
${\bf 5}$ representations. Furthermore, we include a number of GUT
singlets ${\bf 1}$. In order to make contact with the discussion of
subsections \ref{Physics_without_section} and \ref{ss:discrete} we distinguish
representations by an additional $U(1)_1$ charge, corresponding to the
Abelian gauge field $\hat A^1$ introduced above. We indicate the $U(1)_1$ charges of
the ${\bf 10}$, ${\bf 5}$ and ${\bf 1}$ states will by a
subscript $q$ as in
\begin{align}
  {\bf 10}_{ q}\ ,\  {\bf 5}_{ q}\ ,\  {\bf 1}_{ q}\ : \qquad {\bf R}_q \rightarrow  e^{2 \pi i q \Lambda} {\bf R}_q\ ,
\end{align}
where a gauge transformation of $\hat A^1$ acts as $\hat
A^1\rightarrow \hat A^1 + d \Lambda$.

Since we are interested in Yukawa couplings, the relevant terms in the $U(1)$-invariant perturbative superpotential are 
\begin{align}
   W_{\rm pert} : \qquad   \sum_{ q_1+ q_2+ q_3=0}  {\bf 10}_{ q_1} {\bf 10}_{ q_2} {\bf 5}_{ q_3}  \ ,
   \qquad \sum_{ q_1+ q_2+ q_3=0}   {\bf 10}_{ q_1} {\bf \bar 5}_{ q_2} {\bf \bar 5}_{ q_3} \ .
\end{align}
This generically implies that various couplings are absent. As an example, which we will realize in 
F-theory below, let us assume that we have a $4$-split, i.e. $k=4$ in \eqref{eq:split_fund} and \eqref{eq:split_anti} with the representations 
\begin{equation} \label{example_spectrum}
   {\bf 5}_{-6},\ {\bf 5}_{-1},\ {\bf 5}_{4},\ {\bf 10}_{3}, \ {\bf 1}_{5},\  {\bf 1}_{10}\,.
\end{equation}
The perturbatively permitted cubic Yukawas are then
\begin{align}
   \rep{10}_{3} \times \rep{10}_{3} \times \rep{5}_{-6}\,,
    \qquad \rep{10}_{3} \times \bar{\rep{5}}_{1}\times \bar{\rep{5}}_{-4}\ ,
\end{align}  
plus additional couplings involving the singlet states. 

Let us now contrast this to the case in which the $U(1)$ vector field has gained a mass term. As discussed
above, this implies that the low-energy gauge symmetry is reduced to $\mathbb{Z}_2 \times SU(5)$. For our
specific set-up we find that the $\mathbb{Z}_2$ charges are as follows:
\begin{align} \label{eq:Z2_charges}
Q_{\mathbb{Z}_2}(\rep{5}_4) &= 0\,, 
& Q_{\mathbb{Z}_2}(\rep{5}_{-1})&= 1 \,, & 
Q_{\mathbb{Z}_2}(\rep{5}_{-6}) &= 0\,, \nonumber \\
Q_{\mathbb{Z}_2}(\rep{10}_3) &= 1\,, &
Q_{\mathbb{Z}_2}(\rep{1}_5) &= 1\,, &
Q_{\mathbb{Z}_2}(\rep{1}_{10}) &= 0\,
\end{align}
In particular, this means that at masses below the St\"uckelberg mass of our $U(1)$ gauge field,
the two curves $\rep{5}_{4}$ and $\rep{5}_{-6}$ should be indistinguishable. Furthermore, the
singlets $\rep{1}_{10}$ are not charged under any massless gauge field anymore.

Under the remaining gauge symmetry, we expect to find the Yukawa couplings
\begin{align}
\rep{10}_3 \times \rep{10}_3 \times \rep{5}_{-6}\,, \quad
\rep{10}_3 \times \rep{10}_3 \times \rep{5}_4\,, \quad
\rep{10}_3 \times \bar{\rep{5}}_1 \times \bar{\rep{5}}_{-4}\,, \quad
\rep{10}_3 \times \bar{\rep{5}}_1 \times \bar{\rep{5}}_{6}
\end{align}
plus additional couplings involving the singlet states. It is crucial to point out, however, that the coupling
$\rep{10}_3 \times \rep{10}_3 \times \rep{5}_{-1}$ is still ruled out by the $\mathbb{Z}_2$
symmetry and we do not expect it to be realized in our example geometries.

It is particularly interesting to stress the role of the singlets in 
the setup. In the example of section \ref{sec:examples}, we show that the singlet states ${\bf 1}_{10}$
are involved in the Higgsing described in the previous subsection \ref{Physics_without_section}.
In fact, the spectrum \eqref{example_spectrum} arises in the open string interpretation 
of the F-theory setting. The closed string axion appears as the phase of the ${\bf 1}_{10}$ 
using the identification \eqref{open-closed-map}.
Furthermore, we will find in our concrete example that there are couplings of the form
\begin{align}
  \label{eq:singlet-cubic-Yukawa}
  {\bf 1}_{10}  \times {\bf 5}_{-6} \times {\bf \bar 5}_{-4}\, . 
\end{align}
Given such a coupling in the open string picture, one may thus wonder 
whether from the closed string point of view 
a non-perturbative superpotential appears that involves the complex field $G$.
Concretely, inspired by the identification \eqref{open-closed-map} we have in mind terms of the form
\begin{align} \label{non-pert-super}
    W_{\rm non-pert} = \quad \ldots \quad +  \sum_{ q_1+ q_2-r m=0}  A(z) e^{2\pi i r G}  {\bf  5}_{ q_1} {\bf \bar 5}_{q_2}\ .
\end{align}
As we will explain in subsection \ref{Higgstransition}, some of
these couplings are indeed present, and can be reinterpreted in terms
of the classical couplings~\eqref{eq:singlet-cubic-Yukawa}. 

Let us close this subsection with some comments on the non-perturbative couplings \eqref{non-pert-super}.
Superpotential couplings of a similar type induced by stringy instantons 
have been studied intensively in orientifold compactifications as 
reviewed in detail in \cite{Blumenhagen:2009qh}. 
Remarkably, the couplings \eqref{non-pert-super} appear 
to be of somewhat different nature. They do not depend on the K\"ahler moduli and therefore 
are not suppressed at large volume. However, this is not a contradiction to a de-compactification 
argument, since these couplings 
are localized near the intersection of seven-branes. The instantons give a mass for certain ${\bf 5}$-states 
that will therefore be absent in the effective theory for the massless modes only. We will 
see in our concrete examples that this picture is indeed consistent. It would be very 
interesting to perform a more thorough study of the instantons inducing the couplings \eqref{non-pert-super}.
Interestingly, this can already be done in the weak coupling limit.

\subsection{String interpretation of the Higgsing} \label{Higgstransition}

Let us now try to understand better the link between geometric
quantities on the one hand and field theory quantities on the other. We emphasize that the fact
that a new branch of moduli space opens up in the M-theory
compactification, connecting via a geometric transition our Calabi-Yau
background to a large network of spaces, is not essential for our
discussion. An alternative, more self-contained, viewpoint is that we
are studying the physics of the Higgsed (i.e.~deformed) branch close
to a particular point in moduli space where extra degrees of freedom
appear. Nevertheless, we will keep using the M-theory viewpoint for
convenience, since discussions about geometry and M2-brane states can be
easily understood there.

Let us start with the case of the five-dimensional transition, i.e.~a conifold
transition for a Calabi-Yau threefold in M-theory. This case is well
understood by now and we briefly recall the discussion of the
transition given in \cite{Strominger:1995cz,Greene:1995hu}. Take a
Calabi-Yau threefold $\tX$. As we tune some of the complex structure
moduli, there are codimension $R$ subspaces in complex structure
moduli space where $\tX$ develops conifold
singularities. Geometrically, this implies the simultaneous vanishing
of a number of periods
\begin{align} \label{conifold_period}
  z^i = \int_{\Pi_i}\Omega\,, \qquad i=1,\dots,P
\end{align}
with $\Pi_i$ a set of elements of $H^3(\tX,\bZ)$, and $\Omega$ the
holomorphic three-form of $\tX$. More pictorially, we have $P$
3-spheres contracting to zero size. Not all of these 3-spheres are
homologically independent, only $R$ of them are. Our examples all have
$P-R=1$, and henceforth we restrict the discussion to this case for
concreteness.

Consider the defining equation of the Calabi-Yau fourfold without a
section that we will study later, which is of the form\footnote{We
  changed notation with respect to \cite{Anderson:2014yva}, the most
  relevant part of the dictionary for comparison to that paper is
  $\{\tilde{a}_8, \tilde{a}_3, \tilde{a}_7\}\to \{a,f,e\}$.}
\begin{align}
  \label{eq:quartic}
  \begin{split}
    p_{112} & = \tilde{a}_0 w^2 + \tilde{a}_1 y_1^2 w + \tilde{a}_2 y_1
  y_2 w  + \tilde{a}_3 y_2^2 w + \tilde{a}_4 y_1^4\\
  & \phantom{=} + \tilde{a}_5 y_1^3 y_2 + \tilde{a}_6 y_1^2 y_2^2 +
  \tilde{a}_7 y_1 y_2^3 + \tilde{a}_8 y_2^4 \\
  & = 0\,,
  \end{split}
\end{align}
with the $\tilde{a}_i$ being sections of line bundles of appropriate
degree in the base. The conifold locus in moduli space is obtained by
tuning $R$ coefficients in this equation, which allow us to set
$\tilde{a}_8=0$, modulo local coordinate redefinitions. An argument
in~\cite{Anderson:2014yva} then shows that there are conifold
singularities at the $P$ points in the base given by the solutions of
$\tilde{a}_3=\tilde{a}_7=0$.

In the five-dimensional effective field theory, as we approach the
conifold locus, a massive $U(1)$ vector multiplet becomes light. When
we hit the conifold locus in moduli space the massive vector multiplet
becomes massless, and it splits into a massless vector multiplet and a
massless charged hyper. The physics is thus that of an unHiggsing
process. Going in the reverse direction, i.e. taking $\tilde{a}_8\neq
0$, corresponds to giving a VEV to the charged hyper, and thus an
ordinary five-dimensional Higgsing process.

For our purposes it will be useful to understand the geometric
manifestation of this Higgsing in more detail. (The basic picture was
given in \cite{Greene:1996dh}.) Consider the theory at the conifold
locus. We have a massless $U(1)$ vector multiplet\footnote{Typically
  there will be other $U(1)$ vector multiplets in the low energy
  theory, but one can choose a basis in which they decouple from the
  physics of the transition.}, which in M-theory comes from a
supergravity reduction of the form $C_3=A\wedge \omega$, with $A$ the
five-dimensional vector boson and $\omega$ a harmonic two-form in the
threefold $\tX$. By Poincar\'e duality, we can also think of $\omega$
as defining a four-cycle $D$ in $\tX$.

As we start making $\tilde{a}_8\neq 0$, the $U(1)$ should acquire a
mass. The geometric manifestation of this fact is that $\omega$ is no
longer a harmonic form, but rather becomes a low-lying eigenform of
the Laplacian of $\tX$, or dually, the four-cycle $D$ becomes a four-chain
with boundary. In fact, the four-chain is easy to describe: as we deform
away from the conifold locus, the $P$ conifold singularities are
replaced by $P$ three-spheres $S_i$. There is a relation in homology
between these spheres, i.e. there is a four-chain in homology with
boundary on these spheres. This four-cycle is $D$.

Coming back to the $\tilde{a}_8=0$ conifold locus, we have that there
are also $P$ hypermultiplets charged under the $U(1)$. They come from
M2 branes wrapping the vanishing size holomorphic $S^2$ at the
conifold singularity. As we deform away from the conifold locus,
$R=P-1$ hypermultiplets stay massless, and get reinterpreted in the
geometry as complex structure moduli of the $R$ growing classes in
homology, plus the integrals of $C_3$ and $C_6$ over the same homology
classes. The massive vector boson comes from reducing $C_3$ over the
(non-zero) eigenform of the Laplacian connected to the four-cycle
becoming a four-chain in the conifold transition. From this discussion,
it follows that one should identify the closed string axion entering
the St\"uckelberg mechanism in the geometric description of the
massive $U(1)$ given above with the phase of the charged
hypermultiplet getting a VEV and entering the non-linear realization
of the $U(1)$ gauge symmetry becoming massive.

One take-home message from this discussion is that there is a deep
interrelation between the field theory and the geometry, and a duality
dictionary of sorts: what we see in the field theory as a Higgsing of
a field appears in the geometry as a particular four-cycle getting
boundaries and becoming a four-chain. There is also a nice interplay
between field theory and string theory when it comes to the
corrections to the theory: as explained in \cite{Greene:1996dh}, and
further substantiated in \cite{Ooguri:1996me}, in order to reproduce
the right hypermultiplet moduli space metric one expects from field
theory, one should sum an infinite set of non-perturbative corrections
coming from M2 brane instantons in M-theory.

\medskip

A similar picture will hold in the case of compactifications on a
Calabi-Yau fourfold. We now have an M-theory compactification down to
three dimensions, and there is a $U(1)$ symmetry that becomes Higgsed as
we resolve the conifold singularities. The $U(1)$ vector boson comes
from the reduction of $C_3=A\wedge \omega$. Poincar\'e duality now
tells us that we should be looking for a \emph{six}-cycle in the
geometry that opens up in the resolution process and has boundaries
on five-cycles. These five-cycles have a simple interpretation: instead
of having conifold points in the total space, we now have conifold
\emph{curves}. As we deform the defining equation, we obtain a set of
five-cycles given by fibrations of the deformation $S^3$ over the matter
curve being Higgsed.\footnote{Note that this kind of setup has been studied before in
\cite{Intriligator:2012ue}.} The massive $U(1)$ is associated with the
open chain with boundaries on these five-cycles. 
The conifold periods analog to \eqref{conifold_period} can be 
studied using the recent results of \cite{Grimm:2009ef,Bizet:2014uua}. However, 
the relevant couplings, as discussed in subsection \ref{Physics_without_section}, should 
rather be encoded by $J \wedge \Psi$ integrated over the 
five-cycles involved in the transition.

We now obtain a possible reinterpretation of the perturbative field
theory discussion in terms of geometry: the cubic terms that give rise
upon Higgsing to mass couplings between the two $\rep{5}$ curves that
recombine can be understood geometrically as being given by M2
instanton corrections wrapping the contracting three-cycle, as we approach
the conifold point at $\tilde{a}_8=0$. Notice that the discussion is
reminiscent of the $\cN=2$ discussion in
\cite{Greene:1996dh,Ooguri:1996me}. It would be quite interesting
here, for the same reasons, to elucidate the microscopics of the
instanton viewpoint.

\section{A class of elliptic fibrations with discrete symmetries}
\label{sec:examples}

In this section, we present a class of Calabi-Yau manifolds that
realize the effects discussed in the preceding discussion. To do so,
we start in subsection \ref{ss:top} by constructing a class of elliptically fibered manifolds
without section, with fiber a generic quartic in $\bP_{112}$. Next, we
enforce an $SU(5)$ singularity along a divisor of the base manifold
and study the low-energy effective action of F-theory on the
Calabi-Yau manifold.  In section \ref{ss:matter} we find that despite the absence of massless
$U(1)$ gauge factors in the effective action, there are different
matter curves distinguished by a discrete gauge
symmetry that is a remnant of a massive $U(1)$ vector
field.  Furthermore, we encounter that not all the Yukawa couplings
that would naively be allowed by the $SU(5)$ gauge symmetry are
realized geometrically. In fact, we show that those couplings that do
exist correspond precisely to those invariant under the additional
discrete symmetry. 

Moving to the conifold locus in complex structure moduli space we note in 
section \ref{ss:conifold}
that one of the matter curves becomes reducible and splits into two parts.
We note that this is a manifestation of the $U(1)$ becoming massless
at the singular point and the restoration of the full Abelian gauge symmetry.
Resolving the conifold singularities allows us to confirm that the map
between the full $U(1)$ charges and the charge under the discrete remnant group
left over after the Higgsing process is as expected.

\subsection{Hypersurface equation in $\mathbb{P}_{112}$}
\label{ss:top}

Following the discussion in \cite{Anderson:2014yva}, we embed a
genus-one curve inside $\mathbb{P}_{112}$. The most general such
genus-one curve is given by~\eqref{eq:quartic}, which we reproduce
here
\begin{align} \label{e:hypersurface}
  \begin{split}
    p_{112} & = \tilde{a}_0 w^2 + \tilde{a}_1 y_1^2 w + \tilde{a}_2 y_1 y_2 w  + \tilde{a}_3 y_2^2 w
    +\tilde{a}_4 y_1^4 + \tilde{a}_5 y_1^3 y_2 + \tilde{a}_6 y_1^2
    y_2^2 + \tilde{a}_7 y_1 y_2^3 + \tilde{a}_8 y_2^4\\
    & = 0\,,
  \end{split}
\end{align}
where the $\tilde{a}_i$ determine the complex structure of the
genus-one curve. After fibering the curve over a suitable base, the
$\tilde{a}_i$ become sections of line bundles over the base manifold.
As discussed in \cite{Anderson:2014yva}, an elliptic fibration with
such a generic fiber does not have a section, but rather a two-section
defined by $y_1=0$. However, after tuning $\tilde{a}_8 \to 0$ the
genus-one curve becomes singular and the two-section splits into two
independent sections. These can then be most conveniently described
after resolving the singularity obtained by the tuning. Note further
that $\mathbb{P}_{112}$ exhibits an orbifold singularity at the origin
and, in general, this singularity should be resolved. Here, however,
we restrain from doing so and instead impose a condition on
$\tilde{a}_0$ later on that makes sure that our hypersurface does not
hit the orbifold singularity.

Next, let us tune the complex structure coefficients in such a manner
that the elliptic fibration obtains an $SU(5)$ singularity and then
resolve this singularity using methods from toric geometry. In
general, there are many inequivalent ways of creating such a
singularity and then resolving it. Toric resolutions of such
singularities were classified using the formalism of \emph{tops} in
\cite{Bouchard:2003bu} and, for the case of $SU(5)$, evaluated
explicitly in \cite{Braun:2013nqa}. In the language of
\cite{Braun:2013nqa} the ambient fiber space $\mathbb{P}_{112}$
correspond to the polygon $F_4$ and there are three inequivalent
tops\footnote{Put differently, that means that there are three
  different ways of engineering a resolved $SU(5)$ singularity. Note
  that one of the tops, called $\tau_{4,2}$ in \cite{Braun:2013nqa},
  leads to non-minimal singularities even for Calabi-Yau threefolds.}.
Let us pick the first one, called $\tau_{4,1}$ in
\cite{Braun:2013nqa}, and denote the four blow-up variables and the
variable corresponding to the affine node by $e_i$,
$i=0,\dots,4$. Then this choice of $SU(5)$ top implies that the
coefficients $a_i$ must factor according to
\begin{align} \label{e:conditions_top}
 \tilde{a}_0 &= e_0^2 e_1 e_4 \cdot a_0 &
 \tilde{a}_1 & = e_1 e_2 \cdot a_1 &
 \tilde{a}_3 &= e_0 e_3 e_4 \cdot a_3&
 \tilde{a}_4 &=  e_1^3  e_2^4   e_3^2 e_4 \cdot a_4 \nonumber \\
 \tilde{a}_5 &=   e_1^2  e_2^3  e_3^2 e_4 \cdot a_5 &
 \tilde{a}_6 &= e_1 e_2^2 e_3^2 e_4 \cdot a_6 &
 \tilde{a}_7 &=  e_2  e_3^2 e_4 \cdot a_7 &
 \tilde{a}_8 &= e_0  e_2 e_3^3 e_4^2 \cdot a_8\,,
\end{align}
where the $a_i$ are irreducible polynomials and $\tilde{a}_2 = a_2$. Unlike the $\tilde{a}_i$, it is crucial that the 
$a_i$ depend on $e_i$ only through the combination $e_0 e_1 e_2 e_3 e_4$.

\subsection{Non-Abelian matter curves and Yukawa points} \label{ss:matter}
Having tuned the complex structure coefficients in the above manner, the next step is to verify
that this does produce an $SU(5)$ singularity
and to examine what sort of matter representations arise at codimension two in the base manifold.

To do this, let us now compute the Weierstrass form \eqref{eq:Weierstrass}
of the Jacobian of the above genus-one curve.
One finds that the Weierstrass coefficients $f$ and $g$ also depend on the $e_i$ only 
through the combination $e_0 e_1 e_2 e_3 e_4$ and we can therefore go to
a patch in which $e_1=e_2=e_3=e_4=1$ without losing any information.
In that case $f$ and $g$ read
\begin{align}
 f&= -\frac{1}{48} \cdot \Big(a_{2}^{4} - 8 e_0 \cdot  a_{1} \cdot a_{2}^{2} \cdot a_{3} 
 + 8 e_0^2 \cdot (2 a_{1}^{2} a_{3}^{2}  - a_{0} a_{2}^{2} a_{6} + 3 a_{0} a_{1} a_{2} a_{7} ) \nonumber \\
 & \qquad \qquad \ 
 +8  e_0^3 \cdot a_0 \cdot (3  a_{2} a_{3} a_{5} - 2  a_{1} a_{3} a_{6} 
 - 6  a_{1}^{2} a_{8}) \nonumber \\
 & \qquad \qquad \
 + 16 e_0^4 \cdot a_{0} \cdot (- 3  a_{3}^{2} a_{4} + a_{0} a_{6}^{2} -
3 a_{0} a_{5} a_{7}) + 192 e_{0}^{5} \cdot  a_{0}^{2} a_{4} a_{8} 
 \Big)
\end{align}
and
\begin{align}
 g &= \frac{1}{864} \cdot \Big(
 a_{2}^{6}  - 12 e_0 \cdot a_{1} \cdot a_{2}^{4} \cdot a_{3}
 + 12 e_0^2 \cdot a_{2}^{2} \cdot (4 a_{1}^{2}  a_{3}^{2} -  a_{0} a_{2}^{2} a_{6}
 + 3 a_{0} a_{1} a_{2} a_{7}) \nonumber \\
 & \qquad \qquad \
 + 4 e_0^3 \cdot (- 16 a_{1}^{3} a_{3}^{3}
+ 9 a_{0} a_{2}^{3} a_{3} a_{5} + 6 a_{0} a_{1}
a_{2}^{2} a_{3} a_{6} - 36 a_{0} a_{1}^{2} a_{2} a_{3} a_{7}
 - 18 a_{0} a_{1}^{2} a_{2}^{2} a_{8}) \nonumber \\
 & \qquad \qquad \
 + 12 e_0^4 \cdot a_0 \cdot (- 6  a_{2}^{2} a_{3}^{2} a_{4}  - 12 a_{1} a_{2} a_{3}^{2} a_{5}
 + 8  a_{1}^{2} a_{3}^{2} a_{6}  + 4 a_{0} a_{2}^{2} a_{6}^{2} \nonumber \\
& \qquad \qquad \qquad \qquad \qquad - 6 a_{0}^{2} a_{2}^{2} a_{5} a_{7}  - 12 a_{0}^{2} a_{1} a_{2}
a_{6} a_{7} + 18 a_{0}^{2} a_{1}^{2} a_{7}^{2}  +
24 a_{0} a_{1}^{3} a_{3} a_{8} ) \nonumber \\
& \qquad \qquad \
+ 48 e_0^5 \cdot a_0 \cdot  (6  a_{1} a_{3}^{3}
a_{4}  - 3 a_{0} a_{2} a_{3} a_{5} a_{6}  + 2 a_{0} a_{1} a_{3} a_{6}^{2}
+ 18 a_{0} a_{2} a_{3} a_{4} a_{7}  \nonumber \\
& \qquad \qquad \qquad \qquad \qquad \ - 3 a_{0} a_{1} a_{3} a_{5} a_{7} 
- 12 a_{0} a_{2}^{2} a_{4} a_{8} + 18 a_{0} a_{1} a_{2} a_{5} a_{8} 
- 12 a_{0} a_{1}^{2} a_{6} a_{8}) \nonumber \\
& \qquad \qquad \ 
+ 8 e_0^6 \cdot a_0^2 \cdot ( 27 a_{3}^{2} a_{5}^{2}  - 72  a_{3}^{2} a_{4} a_{6}
- 8 a_{0} a_{6}^{3}  + 36 a_{0} a_{5} a_{6} a_{7} - 108 a_{0} a_{4} a_{7}^{2} - 144 a_{1} a_{3} a_{4} a_{8} ) \nonumber \\
& \qquad \qquad \ 
+ 288 e_0^7 \cdot a_0^3 \cdot (-3  a_{5}^{2} a_{8} +8  a_{4} a_{6} a_{8} )
 \Big )\,.
\end{align}
From that it follows directly that the discriminant, defined by $\Delta = 4 f^3 + 27 g^2$, obeys
\begin{align}
 \Delta &= \frac{a_0^2}{16} \cdot \Bigg(
 e_0^5 \cdot  a_{2}^{4} \cdot (- a_{3}
a_{7} + a_{2} a_{8}) \cdot (- a_{2}^{3} a_{4} + a_{1} a_{2}^{2} a_{5} - 
a_{1}^{2} a_{2} a_{6} + a_{1}^{3} a_{7}) \nonumber \\
& \qquad \qquad \ 
- e_0^6 \cdot a_{2}^{2}  \cdot
(a_{2}^{4} a_{3}^{2} a_{4} a_{6} -  a_{1} a_{2}^{3} a_{3}^{2} a_{5}
a_{6} + a_{1}^{2} a_{2}^{2} a_{3}^{2} a_{6}^{2} + 11 a_{1} a_{2}^{3}
a_{3}^{2} a_{4} a_{7} \nonumber \\
& \qquad \qquad \qquad \qquad \quad
- 10 a_{1}^{2} a_{2}^{2} a_{3}^{2} a_{5} a_{7} + 8
a_{1}^{3} a_{2} a_{3}^{2} a_{6} a_{7} - 8 a_{1}^{4} a_{3}^{2} a_{7}^{2}
+ a_{0} a_{2}^{4} a_{4} a_{7}^{2} \nonumber \\
& \qquad \qquad \qquad \qquad  \quad
-  a_{0} a_{1} a_{2}^{3} a_{5}
a_{7}^{2} + a_{0} a_{1}^{2} a_{2}^{2} a_{6} a_{7}^{2} -  a_{0} a_{1}^{3}
a_{2} a_{7}^{3} - 12 a_{1} a_{2}^{4} a_{3} a_{4} a_{8} \nonumber \\
& \qquad \qquad \qquad \qquad  \quad
+ 11 a_{1}^{2} a_{2}^{3} a_{3} a_{5} a_{8} - 10 a_{1}^{3} a_{2}^{2} a_{3} a_{6} a_{8} +
8 a_{1}^{4} a_{2} a_{3} a_{7} a_{8} + a_{1}^{4} a_{2}^{2} a_{8}^{2}) \nonumber \\
& \qquad \qquad \
+ e_0^7 \cdot (a_{2}^{5} a_{3}^{3}
a_{4} a_{5} -  a_{1} a_{2}^{4} a_{3}^{3} a_{5}^{2} + 10 a_{1} a_{2}^{4}
a_{3}^{3} a_{4} a_{6} - 8 a_{1}^{2} a_{2}^{3} a_{3}^{3} a_{5} a_{6} + 8
a_{1}^{3} a_{2}^{2} a_{3}^{3} a_{6}^{2} \nonumber \\
& \qquad \qquad \qquad \quad + 40 a_{1}^{2} a_{2}^{3}
a_{3}^{3} a_{4} a_{7} - 32 a_{1}^{3} a_{2}^{2} a_{3}^{3} a_{5} a_{7} +
a_{0} a_{2}^{5} a_{3} a_{5}^{2} a_{7} + 16 a_{1}^{4} a_{2} a_{3}^{3}
a_{6} a_{7} \nonumber \\
& \qquad \qquad \qquad \quad 
- 12 a_{0} a_{2}^{5} a_{3} a_{4} a_{6} a_{7} + 8 a_{0} a_{1}
a_{2}^{4} a_{3} a_{5} a_{6} a_{7} - 8 a_{0} a_{1}^{2} a_{2}^{3} a_{3}
a_{6}^{2} a_{7} - 16 a_{1}^{5} a_{3}^{3} a_{7}^{2} \nonumber \\
& \qquad \qquad \qquad \quad + 48 a_{0} a_{1}
a_{2}^{4} a_{3} a_{4} a_{7}^{2} - 41 a_{0} a_{1}^{2} a_{2}^{3} a_{3}
a_{5} a_{7}^{2} + 46 a_{0} a_{1}^{3} a_{2}^{2} a_{3} a_{6} a_{7}^{2} \nonumber \\
& \qquad \qquad \qquad \quad
- 36 a_{0} a_{1}^{4} a_{2} a_{3} a_{7}^{3} - 50 a_{1}^{2} a_{2}^{4}
a_{3}^{2} a_{4} a_{8} + 40 a_{1}^{3} a_{2}^{3} a_{3}^{2} a_{5} a_{8} - 
a_{0} a_{2}^{6} a_{5}^{2} a_{8} \nonumber \\
& \qquad \qquad \qquad \quad - 32 a_{1}^{4} a_{2}^{2} a_{3}^{2} a_{6}
a_{8} + 16 a_{0} a_{2}^{6} a_{4} a_{6} a_{8} - 12 a_{0} a_{1} a_{2}^{5}
a_{5} a_{6} a_{8} + 12 a_{0} a_{1}^{2} a_{2}^{4} a_{6}^{2} a_{8} \nonumber \\
& \qquad \qquad \qquad \quad
+ 16 a_{1}^{5} a_{2} a_{3}^{2} a_{7} a_{8} - 40 a_{0} a_{1} a_{2}^{5} a_{4}
a_{7} a_{8} + 34 a_{0} a_{1}^{2} a_{2}^{4} a_{5} a_{7} a_{8} \nonumber \\
& \qquad \qquad \qquad \quad
- 44 a_{0} a_{1}^{3} a_{2}^{3} a_{6} a_{7} a_{8} + 30 a_{0} a_{1}^{4} a_{2}^{2}
a_{7}^{2} a_{8} + 8 a_{1}^{5} a_{2}^{2} a_{3} a_{8}^{2})
+ \mathcal{O}(e_0^8) \Bigg)\,.
\end{align}
Obviously, there is an $SU(5)$ singularity along the GUT divisor %$\{\textrm{pt}\} \times \mathbb{P}^2 \subset \mathbb{P}^1 \times \mathbb{P}^2$
defined by $e_0=0$. Additionally, if $a_0$ has zeros, there will be a further $SU(2)$ singularity whose Cartan divisor
is precisely the divisor obtained from blowing up the $\mathbb{Z}_2$ orbifold singularity of $\mathbb{P}_{112}$. Here
we ignore this additional part by making sure later on that $a_0$ is in fact a constant, which implies that the Calabi-Yau hypersurface
avoids the orbifold singularity.
Furthermore, there are three different curves on the GUT divisor over which the $SU(5)$ singularity is enhanced, namely
\begin{align}
 T & \equiv a_2 = 0 \\
 F_1 & \equiv - a_{2}^{3} a_{4} + a_{1} a_{2}^{2} a_{5} - a_{1}^{2} a_{2} a_{6} + a_{1}^{3} a_{7} = 0 \\
 F_2 & \equiv - a_{3} a_{7} + a_{2} a_{8} = 0\,.
\end{align}
Since we have that 
\begin{align}
 f \rvert_{T=0} &= \mathcal{O}(e_0^2)\,, & g \rvert_{T=0} &= \mathcal{O}(e_0^3)\,, & \Delta\rvert_{T=0} &= \mathcal{O}(e_0^7) \\
 f \rvert_{F_1=0} &= \mathcal{O}(e_0^0)\,, & g \rvert_{F_1=0} &= \mathcal{O}(e_0^0)\,, & \Delta\rvert_{F_1=0} &= \mathcal{O}(e_0^6) \\
 f \rvert_{F_2=0} &= \mathcal{O}(e_0^0)\,, & g \rvert_{F_2=0} &= \mathcal{O}(e_0^0)\,, & \Delta\rvert_{F_2=0} &= \mathcal{O}(e_0^6)
\end{align}
we find that there are $SU(6)$ singularities along the curves $F_i=0$ and that there is an $SO(10)$ singularity at $T=0$. Consequently,
the $F_i=0$ curves host fundamental matter representations, while the $T=0$ curve is the location of the antisymmetric $\rep{10}$
representation of $SU(5)$. We denote the representation located at $F_1=0$ and $F_2$ by $\rep{5}'$ and $\rep{5}''$, respectively.

Before proceeding any further, let us remark here already that without
further gauge symmetries than $SU(5)$, one would not expect to find
different $\rep{5}$-curves as we just have. We therefore expect there
to be an additional gauge symmetry that can differentiate the two
curves. However, from the absence of sections we know that it cannot
be an Abelian gauge group. It will, in fact, turn out to be a discrete
symmetry that distinguishes the $\rep{5}$-curves.

Next, let us consider the Yukawa points on the GUT divisor, i.e. those points at which several of the curves meet
and the singularity is enhanced even further. We first consider points that involve the $\rep{10}$ representation.
Since we have that
\begin{align}
 f\rvert_{T=0} &= -\frac{1}{3} \cdot \Big(
 e_0^2 \cdot  a_{1}^{2}\cdot  a_{3}^{2}
 - e_0^3 \cdot a_0 \cdot a_1 \cdot (   a_{3} a_{6}  + 3  a_{1} a_{8}) \nonumber \\
 & \qquad \qquad \
 + e_0^4 \cdot a_{0} \cdot (- 3  a_{3}^{2} a_{4} + a_{0} a_{6}^{2} -
3 a_{0} a_{5} a_{7}) + 12 e_{0}^{5} \cdot  a_{0}^{2} a_{4} a_{8} 
 \Big) \\
 g\rvert_{T=0} &= \frac{1}{864} \cdot \Big(
 -64 e_0^3 \cdot  a_{1}^{3} \cdot a_{3}^{3}
 +24 e_0^4 \cdot a_0 \cdot a_{1}^{2} \cdot (4   a_{3}^{2} a_{6}  + 9 a_{0}^{2} a_{7}^{2}  +
12 a_{0} a_{1} a_{3} a_{8} ) \nonumber \\
& \qquad \qquad \
+ 48 e_0^5 \cdot a_0 \cdot a_{1} \cdot (6   a_{3}^{3} a_{4}   + 2 a_{0}  a_{3} a_{6}^{2} - 3 a_{0}  a_{3} a_{5} a_{7} 
- 12 a_{0} a_{1} a_{6} a_{8})
+ \mathcal{O}(e_0^6)
 \Big)
\end{align}
we find the enhancements listed in
table~\ref{t:yukawa_non_abelian_deformed}.
\begin{table}[h!]
\begin{center}
\begin{tabular}{cccc|c}
Equation & Involved curves & Singularity & Coupling & Multiplicity\\
\hline
 $\{a_1=0\} \cap \{a_2=0\}$ & $T$, $F_1$ & non-minimal & - & $0$\\
 $\{a_2=0\} \cap \{a_3=0\}$ & $T$, $F_2$ & $E_6$ & $\rep{10} \times \rep{10} \times \rep{5}''$ & $27$\\
 $\{a_2=0\} \cap \{a_7=0\}$ & $T$, $F_1$, $F_2$ & $SO(12)$ & $\rep{10} \times \bar{\rep{5}}' \times \bar{\rep{5}}''$ & $18$
\end{tabular}
\caption{Yukawa couplings involving only non-Abelian representations. Note that all the couplings are located on the GUT divisor
defined by $e_0=0$. The multiplicities were evaluated explicitly for the example manifold given in subsection \ref{ss:example}.}
 \label{t:yukawa_non_abelian_deformed}
\end{center}
\end{table}

Additionally, there are couplings between the two $\rep{5}$-curves and
singlets under the non-Abelian gauge group. We do not give the
explicit equation of the singlet curve here, but note that we find the
couplings list in table \ref{t:yukawa_abelian_deformed}.
\begin{table}[h!]
\begin{center}
\begin{tabular}{ccc|c}
Involved curves & Singularity & Coupling & Multiplicity\\
\hline
  $F_1$, $F_2$ & $SU(7)$ & $\rep{1} \times \rep{5}' \times \bar{\rep{5}}''$ & $108$\\
\end{tabular}
\caption{Yukawa couplings involving both non-Abelian and Abelian representations. Note that all the couplings are located on the GUT divisor
defined by $e_0=0$. The multiplicities were evaluated explicitly for
the example manifold given in subsection \ref{ss:example}.}
 \label{t:yukawa_abelian_deformed}
\end{center}
\end{table}

\subsection{Curve splitting and conifold transition} \label{ss:conifold}

Before going into the details of the particular base we used in order
to compute the precise number of Yukawa points given in the above
tables, let us first, in the spirit of \cite{Anderson:2014yva}, go to
the conifold locus in moduli space, where we obtain a model with two
sections, or equivalently an extra massless $U(1)$. This gives a curve
of conifold singularities located at $a_3 = a_7 = 0$. As noted above,
this corresponds to tuning $a_8 \to 0$.  Interestingly, this
transition has an effect on the $\rep{5}$-curves in the geometry,
since $F_2$ becomes reducible:
\begin{align}
 F_2 \rvert_{a_8=0} = - \underbrace{a_3}_{F_{2,1}} \cdot \underbrace{a_7}_{F_{2,2}}
\end{align}
If we denote the fundamentals at $F_{2,1}=0$ by $\rep{5}''$ and those
at $F_{2,2}=0$ by $\rep{5}'''$ then we find the Yukawa couplings
listed in table \ref{t:yukawa_non_abelian_resolved}.
\begin{table}[h!]
\begin{center}
\begin{tabular}{cccc|c}
Equation & Involved curves & Singularity & Coupling & Multiplicity\\
\hline
 $\{a_1=0\} \cap \{a_2=0\}$ & $T$, $F_1$ & non-minimal & - & $0$\\
 $\{a_2=0\} \cap \{a_3=0\}$ & $T$, $F_{2,1}$ & $E_6$ & $\rep{10} \times \rep{10} \times \rep{5}''$ & $27$\\
 $\{a_2=0\} \cap \{a_7=0\}$ & $T$, $F_1$, $F_{2,2}$ & $SO(12)$ & $\rep{10} \times \bar{\rep{5}}' \times \bar{\rep{5}}'''$ & $18$
\end{tabular}
\caption{Yukawa couplings involving only non-Abelian representations. Note that all the couplings are located on the GUT divisor
defined by $e_0=0$. The multiplicities were evaluated explicitly for the example manifold given in subsection \ref{ss:example}
after transitioning to the conifold point and resolving the singularities appearing there.}
 \label{t:yukawa_non_abelian_resolved}
\end{center}
\end{table}

Furthermore, in table \ref{t:yukawa_abelian_resolved} we summarize the couplings that do not involve the antisymmetric representation.
\begin{table}[h!]
\begin{center}
\begin{tabular}{cccc|c}
Equation & Involved curves & Singularity & Coupling & Multiplicity\\
\hline
  -& $F_1$, $F_{2,1}$ & $SU(7)$ & $\rep{1} \times \rep{5}' \times \bar{\rep{5}}''$ & $54$\\
  - & $F_1$, $F_{2,2}$ & $SU(7)$ & $\rep{1} \times \rep{5}' \times \bar{\rep{5}}'''$ & $54$\\
 $\{a_3=0\} \cap \{a_7=0\}$ & $F_{2,1}$, $F_{2,2}$ & $SU(7)$ & $\rep{1} \times \rep{5}'' \times \bar{\rep{5}}'''$ & $54$\\
\end{tabular}
\end{center}
\caption{Yukawa couplings involving both non-Abelian and Abelian representations. Note that all the couplings are located on the GUT divisor
  defined by $e_0=0$. The multiplicities were evaluated explicitly for the example manifold given in subsection \ref{ss:example}
  after transitioning to the conifold point and resolving the
  singularities appearing there.}
 \label{t:yukawa_abelian_resolved}
\end{table}

We do not give explicit expressions for the singlet curve involved in
the first two couplings, as they are not complete intersections and
contain a large number of terms.

At the conifold locus in complex structure moduli space, we can also
compute the $U(1)$ charges of the matter states using well-known techniques
\cite{Braun:2013yti}. After rescaling the $U(1)$ factor to avoid fractional charges,
we find the following charge assignments:
\begin{align} \label{eq:U1_charges}
\rep{10} &= \rep{10}_{3}\,, & \rep{5'} &= \rep{5}_{-1}\,, &
\rep{5''} &= \rep{5}_{-6}\,, & \rep{5'''} &= \rep{5}_4
\end{align}
Furthermore, we find that the singlet involved in the
 $\rep{1} \times \rep{5''} \times \bar{\rep{5'''}}$ coupling has $U(1)$-charge $10$, while the singlets
 in the other two $\rep{5} \times \bar{\rep{5}}$ couplings have $U(1)$-charge $5$.

\subsection{Discrete charges and forbidden Yukawa couplings}
Finally, let us move away from the conifold locus again by deforming $\tilde{a}_8 \neq 0$.
Looking at the multiplicities of the Yukawa couplings given in tables
\ref{t:yukawa_non_abelian_deformed}, \ref{t:yukawa_abelian_deformed}, \ref{t:yukawa_non_abelian_resolved},
and \ref{t:yukawa_abelian_resolved}, the following picture about the physics of
the deformation process suggests itself rather naturally. The action
takes place on the $\rep{5''}=\rep{5}_{-6}$ and $\rep{5'''} = \rep{5}_4 $ curves, since
they have the same $\mathbb{Z}_2$ charge according to \eqref{eq:Z2_charges}. We observe
that precisely where these two curves intersect, they have a Yukawa
coupling with the $\rep{1}_{10}$ singlet parameterizing the deformation. As this
singlet gets a VEV, the two curves recombine into a single object
that we called $\rep{5}''$ in section \ref{ss:matter}. Since this is a
local operation close to the intersection of the two curves, we expect
the rest of the Yukawa couplings involving the $\rep{1}_5$ singlets to simply come along for the
ride. And indeed, the multiplicities of the Yukawa points are
conserved, if one compares with the results in the previous
section.

To finish this subsection, let us quickly summarize the $\mathbb{Z}_2$ charges of the matter curves
away from the conifold locus. There one finds that\footnote{Note that since we are not at the conifold locus
anymore, $\rep{5''}$ corresponds to the matter curve $F_2=0$.}
\begin{align}
Q_{\mathbb{Z}_2}(\rep{5'}) &=1\,, &
Q_{\mathbb{Z}_2}(\rep{5''}) &=0\,, &
Q_{\mathbb{Z}_2}(\rep{10}) &= 1\,,
\end{align}
which is compatible with the couplings we found in table \ref{t:yukawa_non_abelian_deformed}.
Note that this is precisely what we expect based on the discussion of section \ref{Yukawa-Structures}.
In particular, we find that the coupling
\begin{align}
\rep{10} \times \rep{10} \times \rep{5'}
\end{align}
is not invariant under the $\mathbb{Z}_2$ action and is \emph{not realized geometrically}, although it
would be allowed by all massless continuous symmetries.

\subsection{An explicit example without non-minimal
  singularities}

\label{ss:example}

After keeping much of the previous discussion independent of the
actual choice of base manifold, let us now present the toric data of
an explicit example here. In doing this, it is important to recall
that as soon as one considers three-dimensional base manifolds, there
will generally be non-minimal singularities corresponding to non-flat
points of the fibration \cite{Braun:2013nqa}. We took this into
account in the above discussion, making tables
\ref{t:yukawa_non_abelian_deformed} and
\ref{t:yukawa_non_abelian_resolved} both contain an entry
corresponding to such a non-minimal singularity. The relevant conditions
will generically have non-trivial solutions at codimension three in
the base manifold. The fact that there generically are such non-flat
points does not imply that examples without them are impossible, or
particularly convoluted. The condition one needs to satisfy is
\begin{align}
 \{a_1 = 0 \} \cap \{ a_2 = 0 \} = \emptyset
\end{align}
and as we will now show some simple geometries admit solutions to this
equation.

Our explicit model is as follows. Take a toric ambient space defined
by a fine star triangulation of the rays given in table
\ref{t:example_rays}. As can be seen from the defining data, the
generic ambient fiber space is $\mathbb{P}_{112}$.
\begin{table}[h]
\centering
 \begin{tabular}{ccc|cc|cccc|ccc}
 $u_0$ & $u_1$ & $u_2$ & $v_1$ & $e_0$ & $e_1$ &
 $e_2$ & $e_3$ & $e_4$ & $y_2$ & $y_1$ & $w$ \\
 \hline
 \hline
-3 & 0 & 0 & 0 & 0 & -1 & -2 & -2 & -1
& -1 & -1 & 1 \\
-3 & 0 & 0 & -1 & 0 & 1 & 1 & 0 & 0
& -1 & 1 & 0 \\
0 & 0 & 0 & -1 & 1 & 1 & 1 & 1 & 1 &
0 & 0 & 0 \\
1 & 0 & -1 & 0 & 0 & 0 & 0 & 0 & 0 &
0 & 0 & 0 \\
0 & 1 & -1 & 0 & 0 & 0 & 0 & 0 & 0 &
0 & 0 & 0
 \end{tabular}
\caption{Homogeneous coordinates of the ambient toric space and the corresponding rays of the toric fan.}
\label{t:example_rays}
\end{table}

The base manifold is $\mathbb{P}^1 \times \mathbb{P}^2$ and the
resolved $SU(5)$ singularity discussed in subsection \ref{ss:top} lies
on the base divisor $\{\textrm{pt}\} \times \mathbb{P}^2 \subset
\mathbb{P}^1 \times \mathbb{P}^2$. Note that making the geometric
transition by going to the conifold locus and resolving the conifold
singularities corresponds torically to introducing another ray with
entries $(0, 1, 0, 0, 0)$ as in \cite{Anderson:2014yva}, which
automatically imposes $a_8=0$.

Given the explicit data of the ambient space in which our Calabi-Yau manifold is embedded,
there is an easy way of confirming the absence of non-flat points.
As discussed in \cite{Braun:2013nqa}, at the non-flat points 
one of the irreducible fiber components
grows an extra dimension. In the notation of table \ref{t:example_rays}, the irreducible fiber components
are the horizontal parts of the exceptional divisors $e_i=0$. The irreducible fiber component
which generically jumps in dimension is the one whose ray does not correspond to a vertex of the top, i.e.
$e_4=0$.

Let us therefore examine this component with care. On the divisor $e_4=0$
the hypersurface equation \eqref{e:hypersurface} reduces to
\begin{align}
 p_{112} \rvert_{e_4=0} = \tilde{a}_1 \cdot y_1^2 w + \tilde{a}_2 \cdot y_2^2 w\,.
\end{align}
However, for the above choice of space, one finds that
\begin{align}
 \tilde{a}_1 = e_1 e_2 \cdot \underbrace{\left( \alpha_1 e_0 + \alpha_2 v_1 \right)}_{a_1}\,,
\end{align}
with $\alpha_i$ two generically non-zero constants. In the base, $e_0$ and $v_1$ are just the homogeneous
coordinates of a $\mathbb{P}_1$ and in particular $e_0 = v_1= 0$ is forbidden. As a consequence, there are no
solutions to $e_0 = a_1 = 0$.

\section{Conclusions}

In this paper we studied the physical implications of the presence 
of geometrically massive $U(1)$ gauge fields in F-theory compactifications 
without section. F-theory on a genus-one fibered Calabi-Yau fourfold $\tX_4$
yields a four-dimensional $\mathcal{N}=1$ effective theory that 
can admit an $SU(5)$ GUT group upon engineering appropriate 
singularities of the fibration. We considered the case in which $\tX_4$
does not admit a section, but rather a bi-section. This implies 
that the fourfold cannot be brought into Weierstrass form, but we 
showed that an $SU(5)$ non-Abelian gauge symmetry can be 
explicitly implemented. The absence of a section was argued to correspond 
to the presence of a massive $U(1)$ under which the matter states of the 
GUT are charged. This imposes stringent condition on the allowed 
Yukawa couplings, which we analyzed in detail for a specific example.

We provided two perspectives on the massive $U(1)$ gauge field.
Firstly, we discussed a closed string perspective, where the $U(1)$
becomes massive by 'eating' a closed string axion. This axion arises 
from the R-R or NS-NS two-form in F-theory and the St\"uckelberg 
gauging is dependent purely on the geometry of the seven-brane 
configuration. A dual open string interpretation was given by introducing 
GUT singlets that carry $U(1)$ charge. Geometrically, these singlets 
are most naturally identified at special loci in the complex structure moduli 
space of the Calabi-Yau fourfold at which a curve of conifold singularities is 
generated. At these loci in moduli space the $U(1)$ is massless and 
the spectrum of the four-dimensional theory can be extracted using the 
techniques developed for F-theory compactifications with multiple $U(1)$s 
\cite{Braun:2013yti,Cvetic:2013uta,Borchmann:2013hta}.
Moving away from the singular locus can be interpreted as a Higgsing 
of certain GUT singlets in the open string picture, which corresponds 
to a recombination of seven-branes in F-theory. We also found that 
geometrically a recombination of certain ${\bf 5}$ matter curves occurs in 
this transition. Such behavior is consistent with discrete selection rules imposed by 
the now massive U(1) after integrating them out.

The study of Yukawa couplings has revealed that even when restricting to massless
modes only, the allowed couplings are restricted by discrete selection rules. In the open string 
picture this is due to the well-known fact that after Higgs mechanism only a discrete symmetry 
remains. This also implies that the triple couplings in the superpotential 
involving the Higgsed singlets turn into mass terms, corresponding precisely
to the fact that some of the $\rep{5}$ matter curves recombine in the Higgs branch. Remarkably, 
the closed string interpretation of the couplings involving the Higgsed singlets requires the 
presence of new instanton effects that are not suppressed by a volume modulus. The precise
interpretation of the instanton effects in F-theory or its weak string coupling Type IIB limit 
is still lacking and would be of importance. In M-theory the non-perturbative effects arise 
from M2-branes wrapped on the shrinking 3-spheres along the conifold curve. 
We argued that this geometric picture allows to identify the key ingredients of the field theory 
setup including the massive $U(1)$ arising from the expansion into non-closed forms. 
Clearly, it would be interesting to complete this picture further elucidating the Yukawa 
couplings and their relation to T-branes.

\acknowledgments

We would like to thank Lara Anderson, Volker Braun, and Diego Regalado for illuminating
discussions. I.G.-E. thanks N.~Hasegawa for kind encouragement and
constant support.

\bibliographystyle{JHEP}
\bibliography{refs}

\end{document}